\documentclass[12pt]{article}

\usepackage{epsf}
\usepackage[dvips]{graphicx}
\usepackage{amssymb}

\begin{document}

\begin{center}
{\bfseries COLOR DIAGRAMS FOR NON VACUUM REGGEONS IN HADRON-HADRON
 INTERACTIONS}

\vskip 5mm

V.A. Abramovsky$^{\dag}$, N.V. Radchenko$^{\ddag}$

\vskip 5mm

{\small {\it Novgorod State University }
\\
$\dag$ {\it E-mail: Victor.Abramovsky@novsu.ru }
\\
$\ddag$ {\it E-mail: nvrad@mail.ru }}
\end{center}

\vskip 5mm

\begin{center}
\begin{minipage}{150mm}
\centerline{\bf Abstract} One-to-one correspondence between dual
diagrams of dual resonance model and QCD based color diagrams
describing non vacuum exchanges in $\pi^+\pi^-$, $\pi^{\pm}p$,
$p\bar{p}$ interactions is discussed. Both for dual and color
diagrams there are state with quark-antiquark in $t$ channel and
state, in which only coherent quark string exists, in $s$ channel.
There are no such dual diagrams in $pp$ interaction. Color diagram
for $pp$ interaction was found basing on principle of conformity.
Secondary hadrons spectrum, obtained from this diagram, has
nucleon in its central region. This effect may lead to increase of
baryon chemical potential in nucleus-nucleus collisions in
facilities NICA and FAIR.
\end{minipage}
\end{center}

\vskip 10mm

\section{Introduction}
Processes with multiple production at high energies are
intensively studied both experimentally and theoretically. At that
time behavior of multiple processes at low energies
$\sqrt{s}\approx 2-5$~GeV is known much less. It is believed now
that nucleus-nucleus collisions at low energies may sooner lead to
discovery of phase transition from hadronic matter to the
quark-gluon plasma than nucleus-nucleus collisions at high
energies.

Multiple processes at low energies are dominated by contributions
of non vacuum reggeons. While ``pomeron physics'' is considerably
well explored basing on QCD (\cite{bib1} -- \cite{bib6}, also see
\cite{bib7} and references therein) clear QCD based picture of
multiple processes, associated with non vacuum reggeons is still
missing.

In present work we will study out which ``color diagrams''
correspond to non vacuum reggeons. Analysis of these color
diagrams will show that at low energies baryon number increase in
central region of produced particles spectrum. For nucleus-nucleus
collisions it means that baryon chemical potential is increased in
central region of spectrum. This may lead to discovery of phase
transition to quark-gluon plasma at energies of facilities NICA
and FAIR.

Here we use Okubo basis for $SU(3)$ group. In this basis quark and
antiquark lines are solid colored lines (red, blue and green),
gluon lines are double dotted lines (see Fig.~1). Direction of
color is preserved.

\begin{figure}[!h]
\centerline{
\includegraphics[scale=0.8]{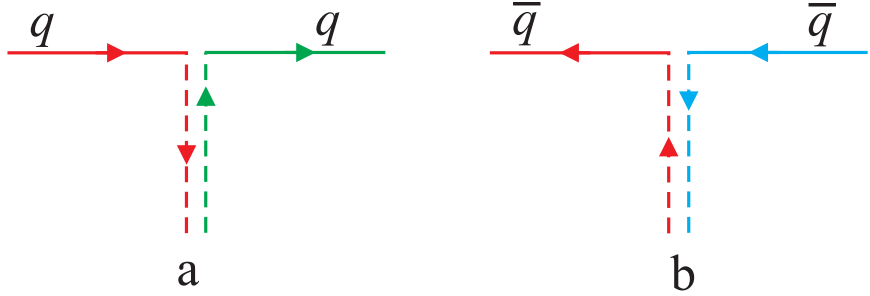}}
\caption{Interaction vertices of quark with gluon (a) and
antiquark with gluon (b).}
\end{figure}

Quark and antiquark lines are continuous, breakup is shown to
indicate direction of color. Gluons are non diagonal and coincide
with gluons in Okubo basis for $U(3)$ group. Diagonal gluons which
are characteristic for $SU(3)$ group will be introduced later.
Summing up all color lines is implied in what follows.

\section{Dually topological diagrams}
Hadrons interactions with exchange of non vacuum reggeon
correspond to soft processes with flavor transfer in $t$ channel.
Since only slow partons softly interact with each other, initial
state configurations where one of valence quarks has low momentum
are very essential.

In frame of dual resonance model (\cite{bib8} -- \cite{bib11})
slowing of quark and reggeon exchange is depicted as dual diagram
in Fig.~2 (we consider $\pi^+\pi^-$ scattering).

\begin{figure}[!h]
\centerline{
\includegraphics[scale=0.8]{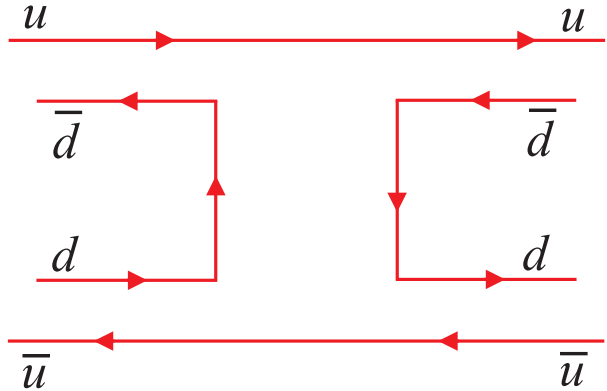}}
\caption{Dual diagram for $\pi^+\pi^-$ scattering}
\end{figure}

In this approach hadrons (mesons) represent string with quark and
antiquark at its endpoints.  When moving in 4-dimensional
space-time string sweeps out 2-dimensional surface. Diagram in
Fig.~2 shows elastic interaction. String endpoints of initial
state merge and further one quark string moves in $s$ channel,
which then splits into two strings. Consequently elastic
scattering amplitude appears, which constitutes smooth
2-dimensional surface. The same 2-dimensional surfaces correspond
to amplitudes of $n$ particles production. More complicated
structure, in which hollow cylinder is glued to poles that are
swept out by strings of initial hadrons, corresponds to pomeron.

Dual resonance model gives independent inference of reggeon
diagram technique, which does not use expansion in colorless
particles of amplitudes or parton wave functions. The AGK
theorem~\cite{bib12} may be obtained also in frame of this
approach~\cite{bib7}. Dual diagrams for $\pi^+p$ and $p\bar{p}$
interactions are given in Fig.~3 and 4.

\begin{figure}[!h]
\centerline{
\includegraphics[scale=0.8]{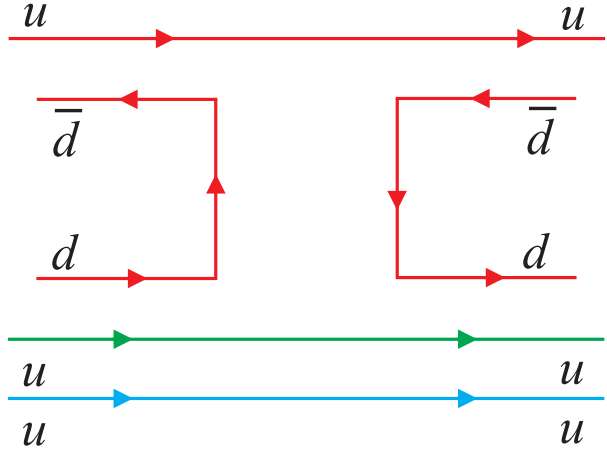}}
\caption{Dual diagrams for $\pi^+p$ interaction. Completely
analogous diagram can be drawn for $\pi^-p$ scattering.}
\end{figure}

In case of $\pi^{\pm}p$ interaction there is stage when only quark
string moves  in $s$ channel, similarly to diagram in Fig.~2.
Endpoints of this string are quark and diquark.

\begin{figure}[!h]
\centerline{
\includegraphics[scale=0.8]{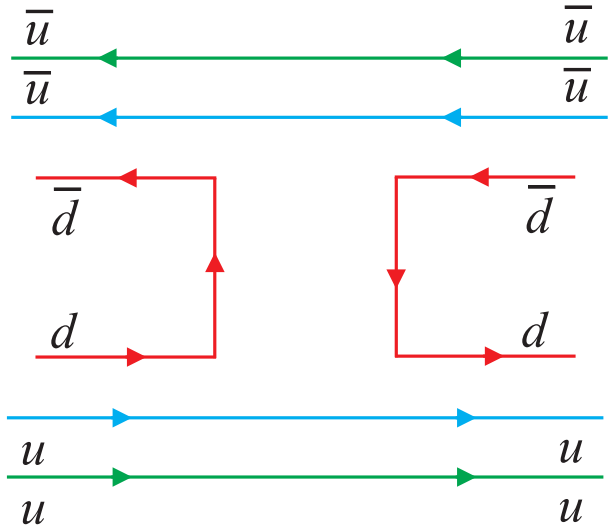}}
\caption{Dual diagram for $p\bar{p}$ scattering.}
\end{figure}

In case of $p\bar{p}$ interaction region is swept out by string
with quark and diquark at its endpoints.

In what follows we will construct color diagrams for non vacuum
reggeons using two results, obtained from consideration of dual
diagrams DRM.

1. In $s$ channel of color diagrams there must exist quark string,
which is not divided into several parts. This string has quarks
(antiquarks) and diquarks (antidiquarks) at its endpoints.

2. In $t$ channel elastic amplitudes, describing reggeon
contributions, must have quark-antiquark pair.

\section{Color diagrams}
We will construct color diagrams for non vacuum reggeons likewise
construction of color diagrams for vacuum exchange~\cite{bib5}.

At first we will consider $\pi^+\pi^-$ scattering (Fig.~5).

\begin{figure}[!h]
\centerline{
\includegraphics[scale=0.8]{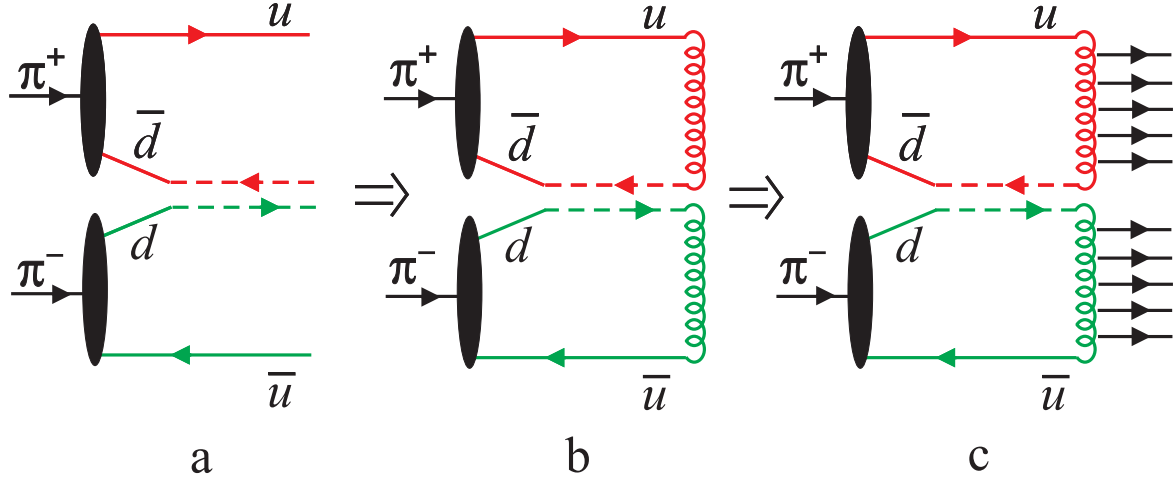}}
\caption{Slow in center-of-mass system quark $d$ and antiquark
$\bar{d}$ annihilate to gluon (a). When quark $u$ and antiquark
$\bar{u}$ move apart, color field string appears (string is shown
as spiral), which is recharged by gluon (b). Then this string
breaks out into secondary hadrons (c).}
\end{figure}

Transfer of color string into secondary hadrons has probability
equal to one, because color objects can not be observed in final
state. Therefore square of diagram module from Fig.~5c is shown in
Fig.~6.

\begin{figure}[!h]
\centerline{
\includegraphics[scale=0.8]{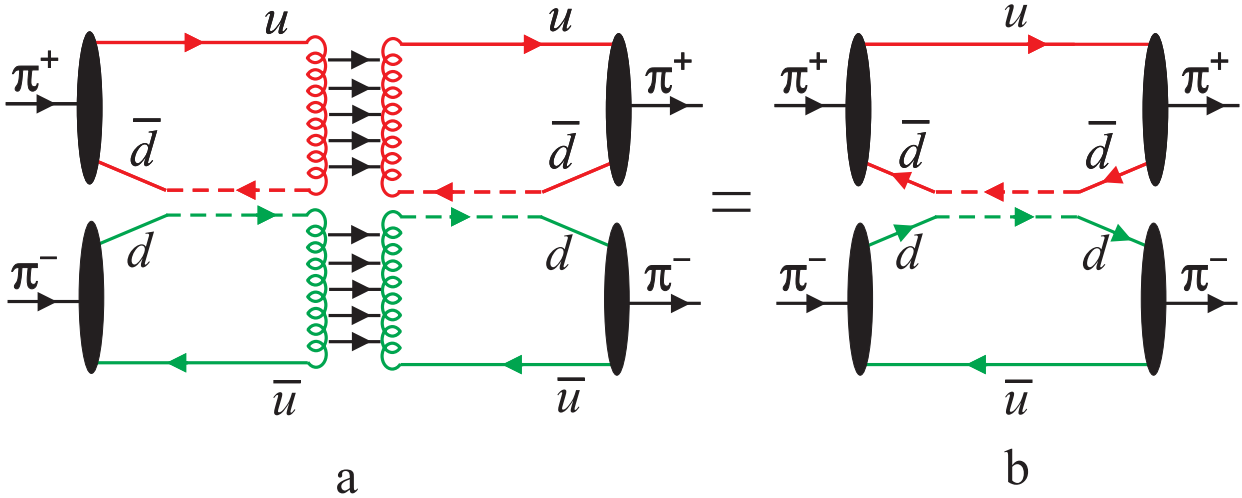}}
\caption{Square of diagram module from Fig.~5c.}
\end{figure}

Transfer from diagram in Fig.~6a to diagram in Fig.~6b is proved
just like transfer to diagram of two gluons exchange
in~\cite{bib5}.

Let us compare diagrams in Fig.~6 with diagram in Fig.~2. Both for
dual diagram in Fig.~2 and color diagrams in Fig.~6 we can state
the following.

1. There is quark string in $s$ channel with quark and antiquark
at its endpoints. Since gluon is pointlike particle, it can only
recharge (recolor) string, but not to split it into two strings
(Fig.~6a).

2. There is quark-antiquark state in $t$ channel (Fig.~6b).

Consequently we have one-to-one correspondence between dual
diagrams in Fig.~2 and color diagrams in Fig.~6. There is the same
one-to-one correspondence  between dual and color diagrams for
$\pi^+p$ and $p\bar{p}$ interactions.

\begin{figure}[!h]
\centerline{
\includegraphics[scale=0.8]{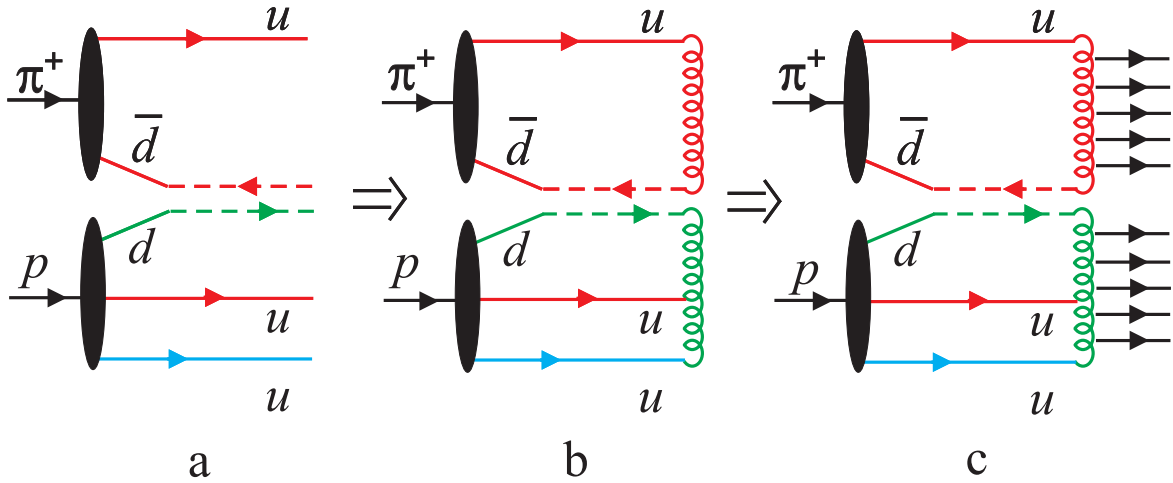}}
\caption{$\pi^+p$ interaction.}
\end{figure}

\begin{figure}[!h]
\centerline{
\includegraphics[scale=0.8]{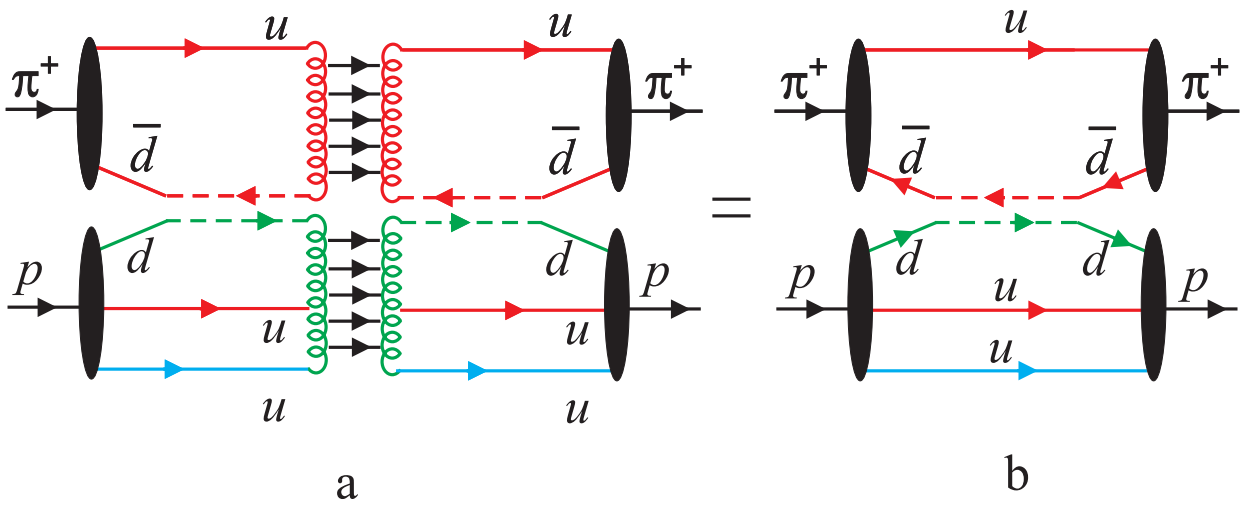}}
\caption{Square of diagram module from Fig.~7c. (a) String in $s$
channel has quark and diquark at its endpoints. (b) Elastic
amplitude in $t$ channel contains quark-antiquark state.}
\end{figure}

\begin{figure}[!h]
\centerline{
\includegraphics[scale=0.8]{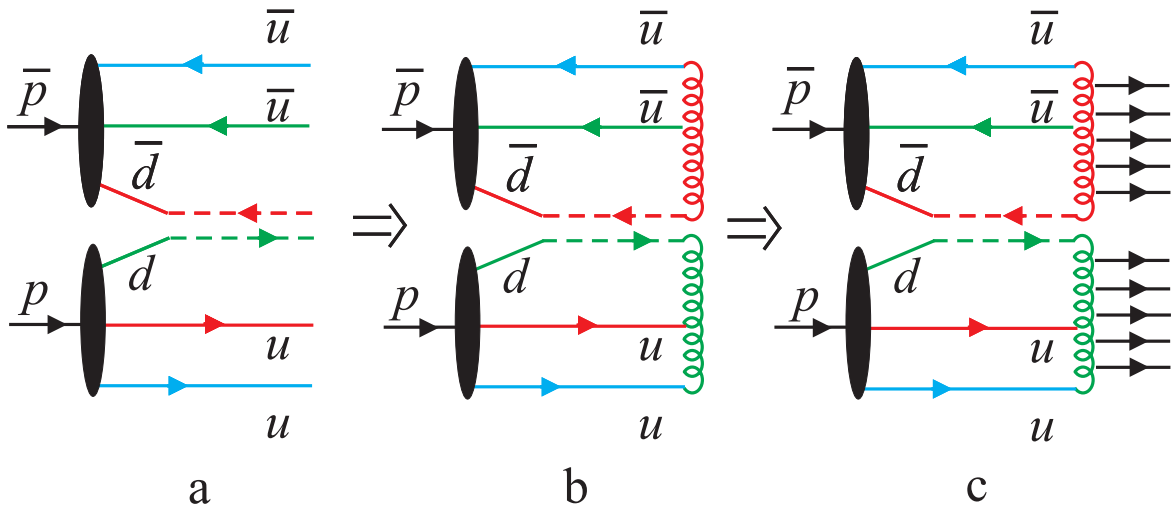}}
\caption{$p\bar{p}$ interaction.}
\end{figure}

\begin{figure}[!h]
\centerline{
\includegraphics[scale=0.7]{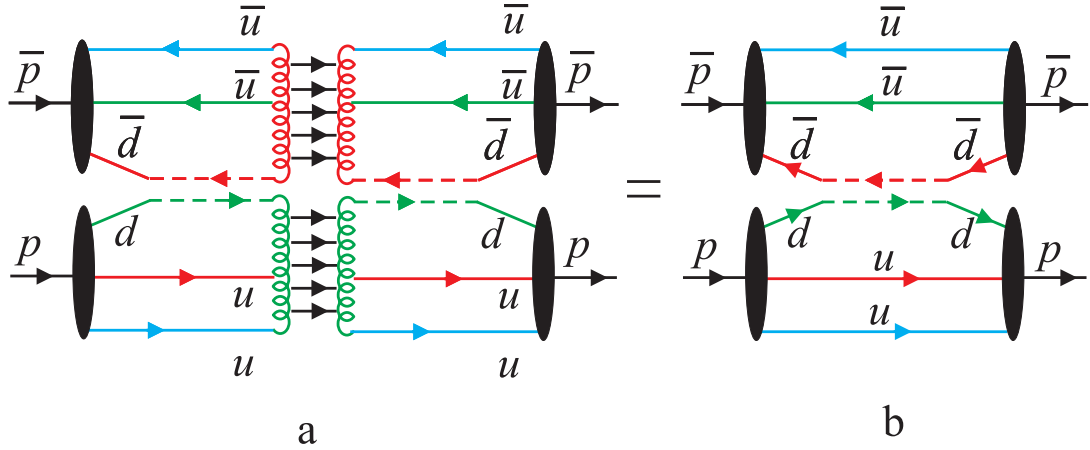}}
\caption{Square of diagram module from Fig.~9c. (a) String
propagates in $s$ channel. Its endpoints are antidiquark and
diquark.  (b)  There is quark-antiquark state in $t$ channel.}
\end{figure}

%\quad\quad\quad
\newpage In both cases of $\pi^+p$ and $p\bar{p}$
scatterings there is one-to-one correspondence between dual and
color diagrams.

\section{Color diagrams for nucleon-nucleon scattering}
There are no dual diagrams for nucleon-nucleon collisions (we will
consider proton-proton scattering as example). This process is
described by so-called twist diagrams, in which scattering of
slowed quarks takes place, but not annihilation of quark and
antiquark. In the first approximation of DRM such diagrams do not
contribute to imaginary part of nucleon-nucleon scattering. So non
vacuum Regge contributions which are present in meson-nucleon and
antinucleon-nucleon interactions must not exist in case of
nucleon-nucleon interactions. But these contributions are visible
in experimental data and parameters of their Regge trajectories
(intercepts and slopes) coincide with  Regge trajectories
parameters of $\pi^\pm p$, $K^\pm p$, $p\bar{p}$, $n\bar{p}$
collisions.

Therefore there must exist a diagram which fulfills requirements
of dual diagrams. We have found two diagrams of process with one
quark string in $s$ channel, Fig.~11.

\begin{figure}[!h]
\centerline{
\includegraphics[scale=0.7]{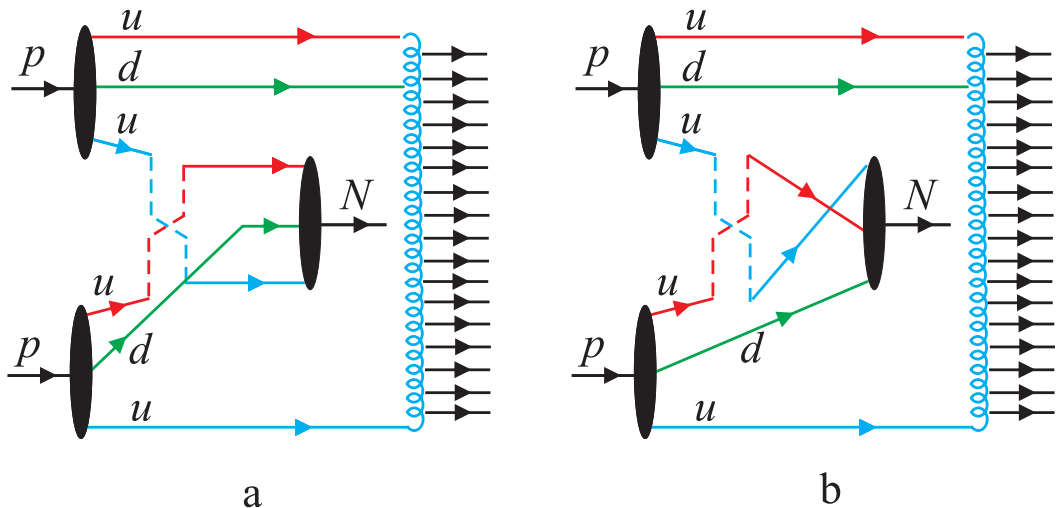}}
\caption{$pp$ interaction.}
\end{figure}

In diagram from Fig.~11a quarks move forward after scattering, in
diagram from Fig.~11b they move backward. In order to form one
quark string in $s$ channel one of protons must be taken in
configuration with slowed diquark. String in $s$ channel has quark
and diquark at its endpoints. Since this string breaks out into
secondary hadrons, then slow state with three quarks also forms
colorless state -- some baryonic resonance.

Configuration with slow quarks in each proton, as shown for
example in Fig.~12, leads to production of two separated in space
quark strings and does not meet the selected principle of
conformity.

\begin{figure}[!h]
\centerline{
\includegraphics[scale=0.8]{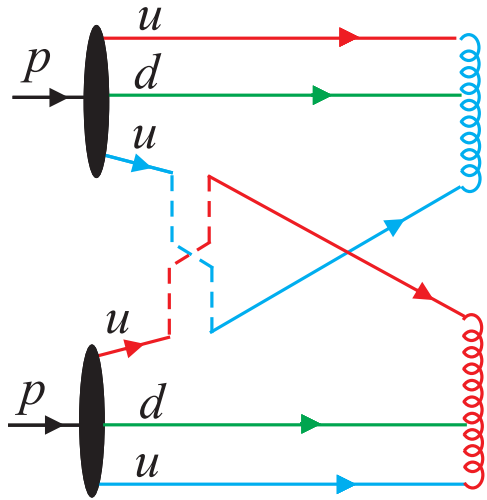}}
\caption{$pp$ interaction.}
\end{figure}

Square of diagram module from Fig.~11a corresponds to two gluons
exchange in $t$ channel and, possibly, it can be described by
trajectory of some $f$-resonances-glueballs. Square of diagram
module from Fig.~11b corresponds in $t$ channel to state 2 quarks
+ 2 antiquarks and is described by low lying non vacuum
trajectory. The only diagram of elastic scattering, which has one
quark string in $s$ channel and quark-antiquark state in $t$
channel, is interference contribution of diagrams from Fig.~11a
and Fig.~11b, it is shown in Fig.~13.

\begin{figure}[!h]
\centerline{
\includegraphics[scale=0.8]{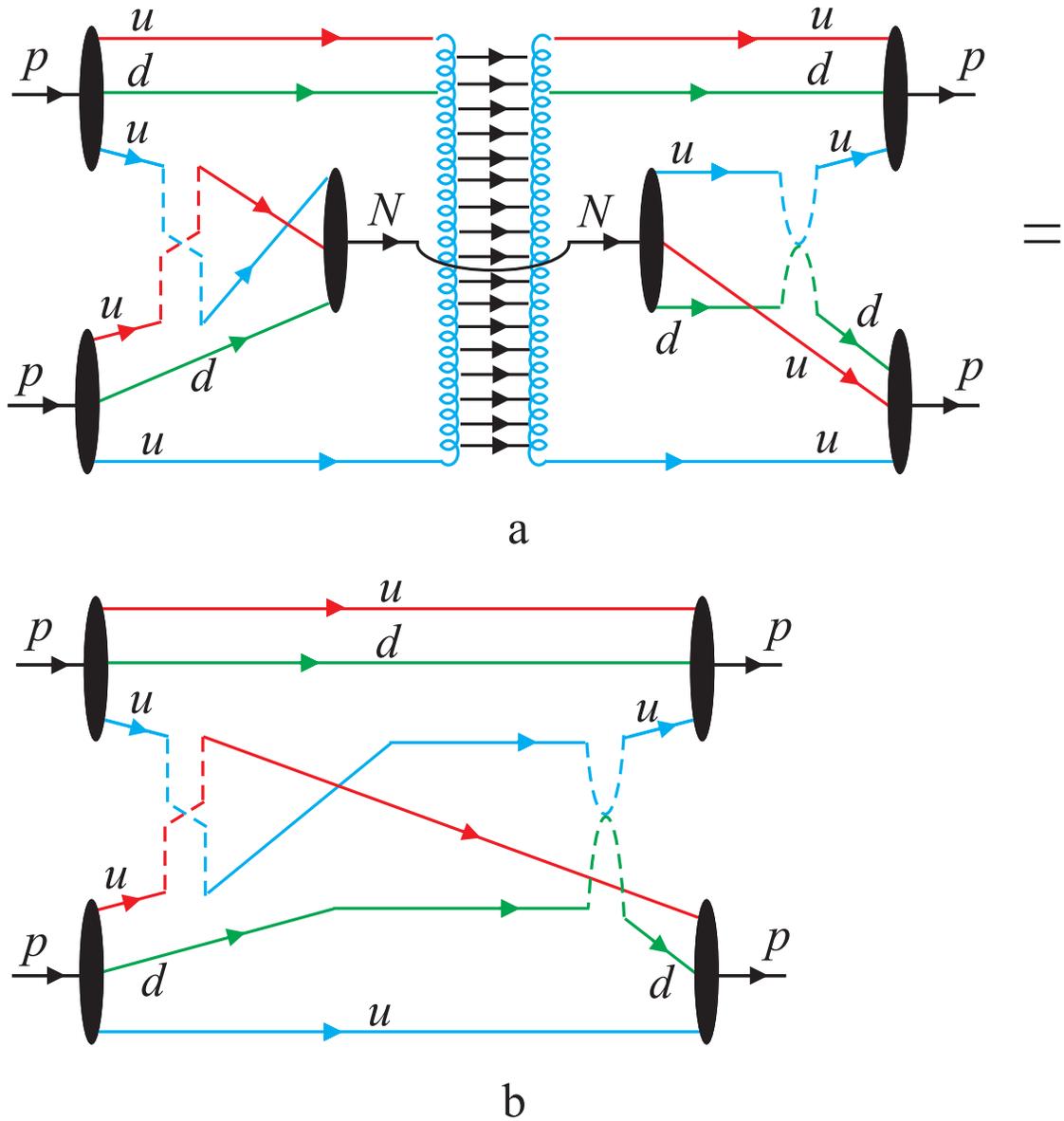}}
\caption{(a) One quark string in $s$ channel. (b) Quark-antiquark
state in $t$ channel.  Diagonal gluons exchange in Okubo basis for
$U(3)$ group is shown in right side.}
\end{figure}

Thus we can argue that leading non vacuum reggeons in
nucleon-nucleon scattering (proton-proton, proton-neutron and
neutron-neutron) result in translocating of baryons from
fragmentation region to central region of secondary hadrons
spectrum.

\section{Conclusion}
The obtained result means that in nucleus-nucleus scattering at
low and intermediate energies baryon number may increase in
central region of secondary hadrons spectrum. This may help in
discovering quark-gluon plasma effects in facilities NICA and
FAIR.

We have come to conclusions by studying structure of color
diagrams for non vacuum reggeons. Evidently, further detailed
analysis is necessary. Though many important results were obtained
only from structure of diagrams in $\lambda\varphi^3$ theory. In
particular, the first proof of the AGK theorem was derived exactly
from analysis of ladder diagram structure in $\lambda\varphi^3$.

One of authors (N.V. Radchenko) gratefully acknowledges financial
support by grant of Ministry of education and science of the
Russian Federation, P1200, and financial support from UNIK NovGU.

\end{document}